\documentclass[aps,prl,reprint,tightenlines,superscriptaddress,amsmath,amssymb]{revtex4-1} 

\usepackage[latin1]{inputenc}
\usepackage{graphicx}
\usepackage{dcolumn}
\usepackage{bm}
\usepackage{bbding}
\usepackage{color}
\usepackage{amsmath}
\usepackage{mathptmx}
\definecolor{red}{rgb}{1,0,0}
  
\newcommand{\tcb}{\textcolor{black}}  
\usepackage{epstopdf}

\usepackage{graphicx}
\usepackage{dcolumn}
\usepackage{bm}
\usepackage{xcolor}
\definecolor{vert}{rgb}{0.41,0.55,0.14}
\newcommand{\be}{\begin{equation}}
\newcommand{\ee}{\end{equation}}

\renewcommand{\Delta}{\triangle}

\newcommand*{\rir}{\right \rangle}
\newcommand*{\lel}{\left \langle}

\newcommand{\B}{{\bm{b}}}
\newcommand{\Bt}{\tilde{{\bm{b}}}}
\renewcommand{\u}{{\bf u}}
\newcommand{\ut}{{\bf \tilde{u}}}
\newcommand{\Bo}{\bm{b_0}}

\newcommand{\Pm}{Pm}

\begin{document}

\preprint{APS/123-QED}

\title{A systematic route to subcritical dynamo branches}

\author{Paul M. Mannix}
\affiliation{Universit\'e C\^ote d'Azur, Inria, CNRS, LJAD, France}

\author{Yannick Ponty}
\affiliation{Universit\'e C\^ote d'Azur, CNRS, Observatoire de la C\^ote d'Azur, Laboratoire Lagrange, France}

\author{Florence Marcotte}
\email{florence.marcotte@inria.fr}
\affiliation{Universit\'e C\^ote d'Azur, Inria, CNRS, LJAD, France}

\begin{abstract}
We demonstrate that the nonlinear optimisation of a finite-amplitude disturbance over a freely evolving and possibly even turbulent flow, can successfully identify subcritical dynamo branches as well as the structure and amplitude of their critical perturbations. As this approach does not require prior knowledge of the magnetic field amplification mechanisms, it opens a new avenue for systematically probing subcritical dynamo flows.
\end{abstract}

\maketitle

Long-lived astrophysical magnetic fields display a remarkable diversity of spatial scales, structures and intensities - with detected fields of order $\sim 1\mu$G in galaxies \cite{Kulsrud2008}, to $\sim 10^{14}$G in magnetars \cite{Vaspi2017}. A century ago, Joseph Larmor argued that intense magnetic fields can be born out of a dynamo instability \cite{Larmor1919}, whereby favorable flow motions amplify a magnetic seed field in the electrically conducting fluid layers of celestial bodies. Ever since, the origin of astrophysical magnetic fields has continued to raise many fundamental questions.

Indeed, exhibiting a flow capable of amplifying and maintaining a magnetic field by dynamo action has proved a challenging task, both from a theoretical and experimental point of view \tcb{\cite{Rincon2019}}
: firstly, fluid motions need to be vigorous enough, and the fluid resistivity sufficiently low, for induction to overcome the destructive effect of ohmic dissipation. As a result, igniting a dynamo instability out of a weak magnetic seed, often translates into intractable parameter regimes for global numerical simulations or unsustainable energy costs for laboratory \tcb{devices}, so that experimental dynamos are very scarce, \tcb{whether in liquid metals \cite{Gailitis2018,Muller2004,VKS} or plasma flows \cite{Tzeferacos2018,Bott2021}}. Secondly, flows too simple - in the sense that they present too many symmetries, such as (typically) Keplerian flows - are linearly stable to dynamo action \cite{Moffatt1978}.

In some situations however, a dynamo can be triggered by \textit{finite-amplitude} disturbances even in a highly symmetric or (comparatively) poorly conducting flow. This springs from the nonlinear nature of the magnetohydrodynamics (MHD) equations, and can occur whenever the feedback of the magnetic field on the flow, 
sustains its own amplification by subcritical dynamo instability. Natural systems where subcritical dynamos are assumed to operate abound, with important examples including the Earth's core \cite{Roberts1978}, but also accretion disks and radiative stars \cite{Rincon2007,Spruit2002}, where magnetogenesis is poorly understood despite its suspected role in explaining the observed anomalous transport of angular momentum \cite{Balbus2003,Aerts2019}.

Due to the lack of a systematic nonlinear stability method however, the identification and modeling of subcritical dynamos so far involves a substantial amount of luck. Successful attempts have relied on imposing (and subsequently removing) a specifically tailored electromotive force or magnetic field \cite{Rincon2007,Guseva2017}. This however, requires some prior knowledge of the field's structure and the intensity required to sustain a dynamo in a given flow. Another way is identifying a dynamo branch as it linearly bifurcates from the base state, and then following the branch backwards in parameter space, below its linear instability threshold - a method however doomed to fail when the threshold parameter regime is intractable, or when the base state is always linearly stable to dynamo instability.

In this Letter, we show how subcritical dynamo branches can be systematically explored in a given MHD base flow. Furthermore, we aim at determining the \textit{minimal dynamo seed} - that is, the spatial structure of the smallest-amplitude (and often low dimensional) magnetic disturbance, capable of nonlinearly triggering a self-sustaining dynamo in a given system. Our approach builds on the mathematical tools of optimal control and nonmodal stability analysis \cite{Peter2007}, which have recently been used in the context of subcritical transition to hydrodynamic turbulence in shear flows \cite{Pringle2010,Eaves2015,Kerswell2018}.
The robustness of the proposed method is demonstrated by considering two contrasting reference flows - namely, a local model of laminar quasi-Keplerian Couette flow
, and a \tcb{turbulent} Taylor-Green flow.

To that end, we consider the MHD system:
\begin{eqnarray}
\frac{\partial \u}{\partial t} + (\u \cdot \nabla) \u - \nu \Delta \u  + \nabla P - (\nabla \times \B ) \times \B && \ = \ 0, \label{eqn:NS} \\
\vspace{-0.5ex}
\frac{\partial \B}{\partial t} -  \nabla \times \big( \u \times \B \big) - \eta \Delta \B && \ = \ 0, \label{eqn:ind} \\
\qquad \qquad \nabla \cdot \u &&\ = \ 0,\\
 \qquad \qquad \nabla \cdot \B && \ =\ 0,
\label{eqn:divB}
\end{eqnarray}
where $\u(\bm{x},t)$, $\B(\bm{x},t)$ and $P(\bm{x},t)$ are respectively the velocity, magnetic and pressure fields; $\nu$ is the kinematic viscosity and $\eta$ the magnetic diffusivity. Here the magnetic field is rescaled by $1/\sqrt{\rho \mu_0}$ as an Alfv\`en velocity, where $\rho$ is the (constant) fluid density and $\mu_0$ the void magnetic permeability.

\begin{figure*}
\begin{center}
\hspace{-5ex}
\includegraphics[width=\textwidth,clip=true]{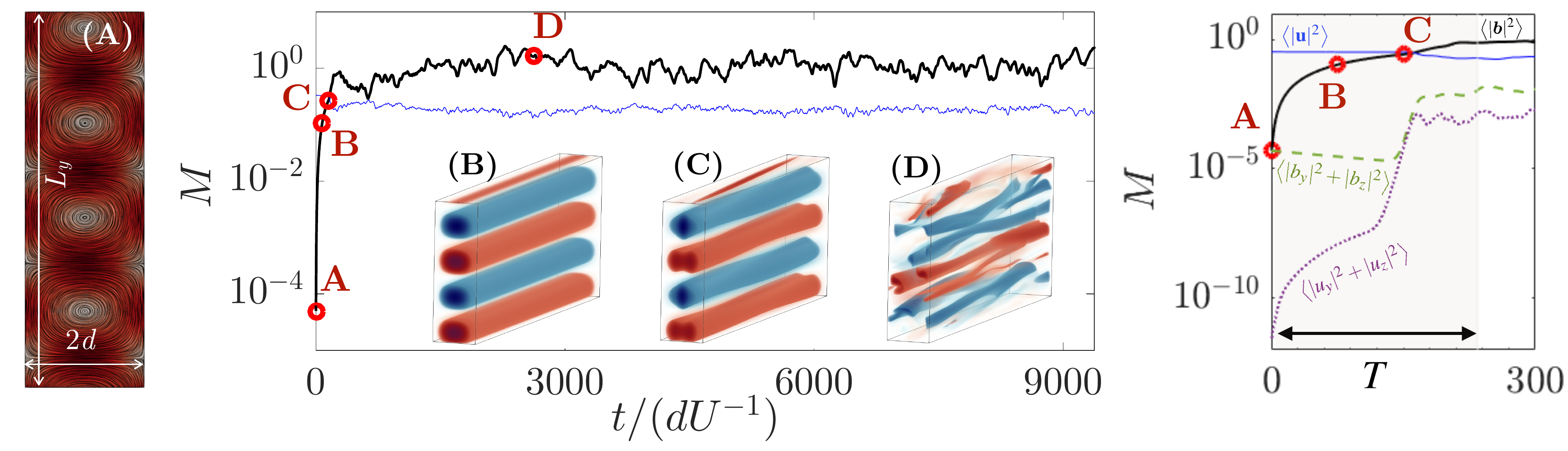}
\caption{\small{Time-series of the magnetic energy (thicker line) and kinetic energy (thinner line) densities, from DNS of the quasi-Keplerian Couette flow at $Re=25$ and $Pm=75$ (resolution: $N_x,N_y,N_z = \{96,192,96\}$). A blowup on the optimisation window \textit{(Right)} shows the same quantities, along with the energies of the transverse fields.
Letters mark the times of the various snapshots. The structure of the optimal seed (A) is highlighted by magnetic streamlines in transverse cut \textit{(Left)}. The snapshots in inset show the amplitude of the streamwise component $b_x$, illustrating the $\Omega$-effect (B), destabilization of the large-scale field to MRI (C) and fully 3D saturated dynamo (D). \tcb{(Note that the MRI has already saturated by the optimization time T.)}}}
\label{fig:Couette}
\end{center}
\end{figure*}

At long times, any initial condition $\Bo=\B(\bm{x},0)$ capable of nonlinearly exciting a dynamo instability, will yield strong amplification of the magnetic energy. 
We thus consider the following question: given a reference flow $\bm{u_0}=\bm{u}(\bm{x},0)$, and an initial magnetic energy density budget $M_0=\lel |\Bo|^2 \rir$ (where $\lel f \rir = \tfrac1V \int_V f dV$ denotes spatial average), what initial condition $\Bo$ maximises the objective function $J(\Bo)=\int_0^T \langle |\B|^2 \rangle dt$ by some target time $T$? 
Maximisation of the magnetic energy has been previously used to optimize a frozen velocity field in the (linear) kinematic dynamo problem \cite{Willis2012, Chen2015,Herreman2016,Herreman2018}.
Here we address the optimisation question by extremising the following \tcb{functional}:
\begin{eqnarray}
\mathcal{L} =  \int_0^T&& \langle |\B|^2 \rangle dt \;  - \int_0^T \langle \Bt \cdot \bigg[ IND  \bigg] \rangle dt \;  - \int_0^T  \langle \tilde{\Pi} \ ( \nabla \cdot \B ) \rangle dt \nonumber \\
&& - \int_0^T \langle \ut \cdot \bigg[ NS \bigg] \rangle dt \; - \int_0^T \langle \tilde{P} \ (\nabla \cdot \u ) \rangle dt,
\label{eqn:Opt_Functional}
\end{eqnarray}
where $\ut(\bm{x},t)$, $\Bt(\bm{x},t)$, $\tilde{P}(\bm{x},t)$, $\tilde{\Pi}(\bm{x},t)$ are Lagrange multipliers used to enforce the nonlinear constraints \eqref{eqn:NS}-\eqref{eqn:divB} in the whole domain for $t \in [0,T]$, and "NS", "IND" stand for the left-hand side (LHS) of \eqref{eqn:NS} and \eqref{eqn:ind}, respectively.
Maximisation with respect to $\Bo$ requires estimation of the variational derivative $\delta \mathcal{L}/\delta \Bo$, which has to vanish when an optimum is reached. \tcb{This estimate is obtained at affordable computational cost using a direct-adjoint looping method. First, canceling the variational derivatives of $\mathcal{L}$ provides (i) compatibility conditions that relate the physical (``direct'') variables $\u, \B$ to the Lagrange multipliers (or ``adjoint variables'') $\ut, \Bt$ at time T; (ii) backward evolution equations and boundary conditions for the adjoint variables,}  and (iii) an optimality condition relating the desired information $\delta \mathcal{L}/\delta \Bo$ to the adjoint fields \tcb{at $t=0$}, which in the present case reduces to
\be
\frac{\delta \mathcal{L} }{\delta \B_0} = \Bt(\bm{x},0) + \nabla \tilde{\Pi}.
\label{eqn:adjIC}
\ee
\tcb{Then for a given reference flow $\u_0$, the looping procedure uses (i)-(iii) as follows}: starting from a (random) first guess $\Bo$ with energy $M_0$, we evolve $\u, \B$ from $t = 0 \to T$ according to \eqref{eqn:NS}-\eqref{eqn:divB}. We then \tcb{use condition (i), here that $\ut(T)=\Bt(T)=0$,} to evolve $\ut, \Bt$ back from $t = T \to 0$ according to the adjoint equations \tcb{(ii):}
\begin{eqnarray}
- \frac{\partial \ut }{\partial t} &&= N(\ut,\u) + \B \times \big( \nabla \times \Bt \big) + \nu \Delta \ut +\nabla \tilde{P},\label{eqn:adj_NS}\\
 - \frac{\partial \Bt }{\partial t} &&=  -N(\ut,\B) \tcb{- \u \times \big(\nabla \times \Bt \big)} + \eta \Delta \Bt + \nabla \tilde{\Pi} + 2 \B,
\label{eqn:adj_ind}
\end{eqnarray}
\tcb{where $N(\ut,\u) \equiv ( \nabla \times \u ) \times \ut  + \nabla \times (\ut \times \u)$, and $\ut$ and $\Bt$ are subject to divergence-free conditions \cite{Guseva2015} and the same boundary conditions as $\u,\B$, respectively.}
This yields the desired gradient information \eqref{eqn:adjIC} at $t=0$, which we use to update $\B_0$ by means of a classical descent method. The latter is combined with the rotation method of \cite{Douglas1998,Foures2013} to enforce the condition that $\B_0$ be of energy $M_0$. The whole process is then repeated until the system converges to an optimum. The resulting optimal seed $\B_0$ exhibits whatever spatial structure yields largest amplification of the magnetic energy by target time $T$. In order to rule out transient states, \tcb{or even self-killing dynamo processes \cite{Fuchs1999,Miralles2015,Cebron2019}}, the optimal seed is then used as an initial condition for integrating the governing equations \eqref{eqn:NS}-\eqref{eqn:divB} (direct numerical simulation (DNS)) over longer timescales $t \gg T$, which finally confirms whether a self-sustained dynamo develops or not.

\begin{figure*}
\begin{center}
\hspace{-5ex}
\includegraphics[width=\textwidth,clip=true]{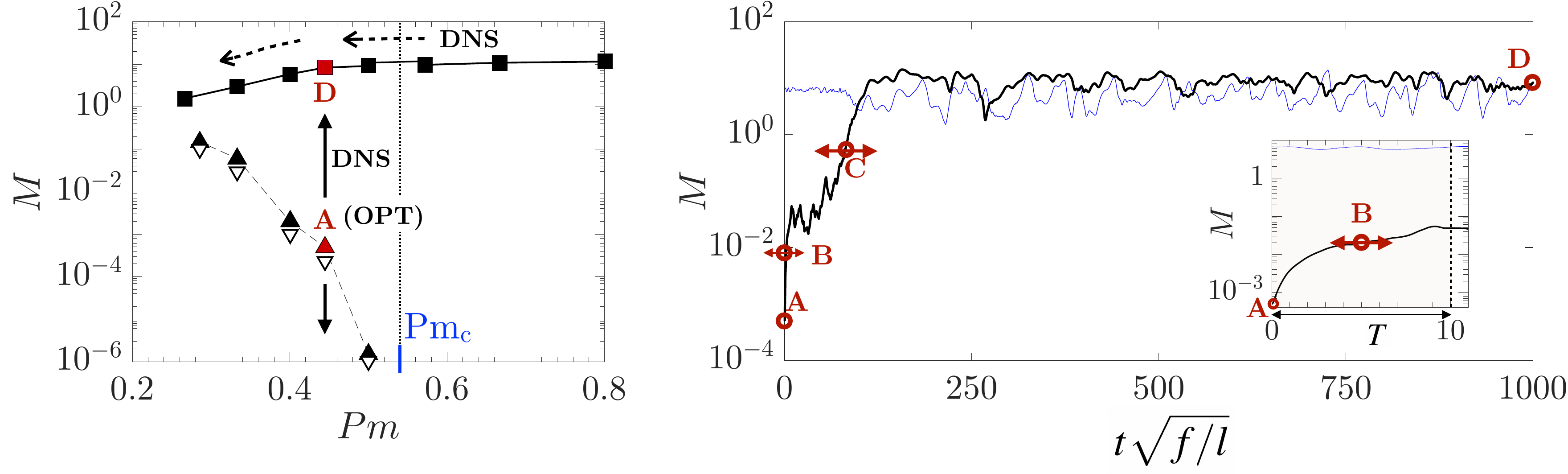}
\caption{\small{\textit{Left:} bifurcation diagram of the Taylor-Green dynamo, for $Gr=1875$. \tcb{Magnetic energy of ($\blacksquare$) the saturated dynamo states  and ($\blacktriangle$) the smallest magnetic disturbances that trigger a subcritical dynamo, as found by nonlinear optimisation with $T=10\sqrt{l/f}$. $\triangledown$: failed dynamo seeds.} \textit{Right:} Typical time-series of the kinetic energy (thinner line) and magnetic energy (thicker line) densities, as the minimal seed (A) spontaneously evolves toward the saturated dynamo state (D). Letters mark the times of the snapshots shown in Fig.\ref{fig:TG2}.}}
\label{fig:TG1}
\end{center}
\end{figure*}

We first consider a laminar incompressible flow, sheared between two infinite parallel rigid plates, in a rotating frame of reference (rotating plane Couette flow). This highly symmetric flow is \tcb{linearly stable to dynamo action} by virtue of Zeldovich's antidynamo theorem \cite{Zeldovich1957}. The existence of subcritical dynamo solutions in quasi-Keplerian Couette flow was numerically demonstrated in \cite{Rincon2007}: 
using Newton iteration they construct steady dynamo states, which we find to be stable however, only with certain symmetries enforced. Their approach relies on prescribing an artificial electromotive force, carefully chosen so as to replenish a large-scale field prone to (linear) destabilization via the magnetorotational instability (MRI). Then by progressively removing it until nonlinear interactions between MRI modes can take over in replenishing the large-scale field, they close the dynamo loop. Because the obtained solutions are time-independent, their approach is inherently restricted to low Reynolds numbers. 

Here we choose the same geometry and boundary conditions as described in \cite{Rincon2007}. The domain aspect ratios are $L_x/L_z=\pi/0.375$ and $L_y/L_z=\pi$ (where $L_x$, $L_y$ and $L_z=2d$ are the domain size in the streamwise, spanwise and shearwise directions, respectively). The perfectly conducting plates are located at $z=-d$ and $z=d$. They countermove shearwise at velocity $+U$ and $-U$, respectively, in the frame of reference rotating about the spanwise direction at constant angular speed $\Omega=2U/3d$. The flow regime is described by the fixed magnetic Prandtl number $\Pm=\nu/\eta=75$ and a somewhat higher Reynolds number than in \cite{Rincon2007}, $Re=U d/\nu=25$.

The forward 
and adjoint 
equations (with additional terms $2\Omega \times \u$ and $-2 \Omega \times \ut$ in the LHS of \eqref{eqn:NS} and \eqref{eqn:adj_NS} respectively to account for Coriolis acceleration) are solved with the pseudo-spectral code Dedalus \cite{Burns2020}, \tcb{using a Chebyshev spectral decomposition in the shearwise direction, Fourier decomposition in $x$ and $y$, and a second-order modified Crank Nicolson Adams Bashforth time-stepping scheme \cite{Ascher1995}}. Our implementation of the forward equations 
has been verified against published results \cite{Rincon2007,Rincon2007Hydro} and our adjoint has been carefully tested to ensure a consistent gradient. For optimisation runs, we use a numerical resolution of \tcb{$N_x,N_y,N_z = 54,144,72$} gridpoints; all results presented here are robust to changes in spatial and 
temporal resolution. The target time for optimisation corresponds to $T=234 d/U$ (or equivalently $T=T_\eta/8$, where $T_\eta=d^2 /\eta$ is the ohmic timescale). Importantly, \textit{no assumptions are made} on the symmetries or time-dependence of a dynamo solution, if there exists one at such $Re$. Keeping all flow parameters fixed and applying the procedure described above for some random energy budget $M_0$, we then decrease or increase $M_0$ depending on whether a dynamo has been found or not. Convergence of the optimisation runs is assessed both in terms of the saturation of the objective function and the significant decrease of a gradient residual defined as in \cite{Foures2013} (eq. C2),
 for which a reduction of at least 2 or 3 orders of magnitude is achieved.
 
Figure \ref{fig:Couette} shows the time-series of the spatially integrated kinetic and magnetic energies, obtained using DNS from a typical optimal seed field $\Bo$ (identified here for $M_0=5\cdot 10^{-5}$). These show that a sustained, fluctuating dynamo state can be reached, with saturation energy $\sim 10^5$ times larger than that of the seed field. The snapshots in the inset show the typical development of the MRI-driven dynamo instability, from seed to nonlinear saturation. Dominated by its transverse components, the nearly streamwise invariant seed structure (A), efficiently exploits the base flow's shear to yield rapid generation of a large-scale streamwise magnetic field (B), through a process equivalent to the so-called $\Omega$-effect \cite{Moffatt1978}. Upon exceeding a critical amplitude, this large-scale component destabilizes and the nonlinear interactions of the resulting fully three-dimensional states (C) close the dynamo loop. In contrast to \cite{Rincon2007}, the obtained dynamo solution is highly fluctuating but stable. \tcb{Note that, while we chose here a large $Pm$ following \cite{Rincon2007}, the same approach can be used in the future to investigate smaller-$Pm$ regimes more relevant for accretion discs \cite{Balbus2008} or stellar interiors \cite{Spruit2002}.}


\begin{figure*}
\begin{center}
\hspace{-5ex}
\includegraphics[width=\textwidth,clip=true]{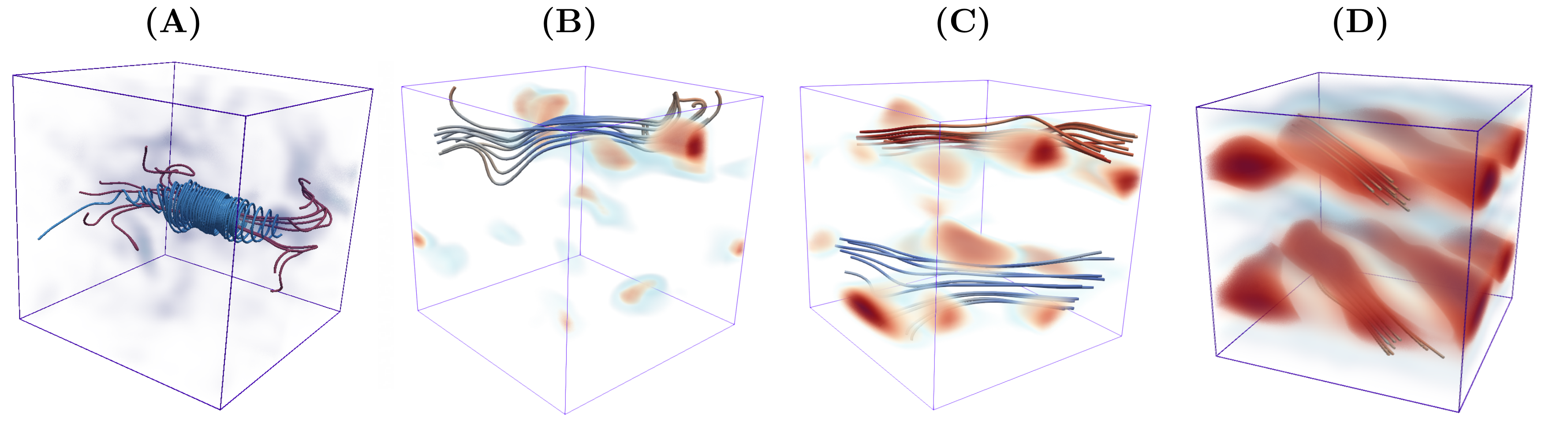}
\caption{\small{\textit{From Left to Right:} Snapshots of (A) the minimal seed for the Taylor-Green dynamo at $\{Gr=1875;Pm=0.44\}$ (magnetic lines in blue, current lines in red), (B) magnetic field during the algebraic growth (rescaled and time-averaged over the growth period); (C) growing magnetic field during the exponential growth (rescaled and time-averaged), (D) saturated state (recovers \cite{Yannick2007}). The volume rendering shows the density of magnetic energy. The magnetic energy at saturation displays cigar-shaped structures aligned in two parallel planes.}}
\label{fig:TG2}
\end{center}
\end{figure*}

To demonstrate the robustness of our approach, we now consider a strongly fluctuating Taylor-Green flow in a triply periodic cube of size $L$. This vortical flow is driven by a constant forcing, introduced in (\ref{eqn:NS}) through an additional forcing term
as per \cite{Yannick2007}. 
Taylor-Green flow is known to be prone to linear dynamo instability, and due to its symmetries has been studied as a simple model for the Von K\'arm\'an dynamo experiment. 
Using continuation from the linearly unstable regime, \cite{Yannick2007} have shown that this flow harbors a subcritical dynamo, in a region of $\{Re,Pm\}$ space, where no infinitesimal seed fields can be amplified. Our aim is to test: if our approach directly identifies this subcritical branch, \tcb{despite velocity fluctuations}, and if the minimal seed capable of triggering the dynamo, bears any topological resemblance with the saturated state.
The pseudo-spectral code {\sc cubby} \cite{Yannick2005} is used for solving the direct and adjoint equations.

The control parameters are $\Pm$ and the Grashof number $Gr=f \, l^3/\nu^2=1875$, where $l=L/(2\pi)$ is the unit length and $f$ the intensity of the Taylor-Green forcing. \tcb{(A diagnostic Reynolds number built on the root-mean square velocity is computed as in \cite{Yannick2008}, with $Re_{rms} = 194.7$.)} For each value of $\Pm$ considered here, we use the target time $T=10 \sqrt{l/f}$, and apply the procedure described above for some (large) energy budget $M_0$, accompanied by DNS for $\sim 100 T$. We typically use a numerical resolution of $64^3$ gridpoints.

The black squares in the bifurcation diagram of Fig.\ref{fig:TG1}\textit{(Left)} denote the dynamo branch previously identified by \cite{Yannick2007},
while $\Pm_c \sim 0.54$ denotes the linear dynamo instability threshold, above which infinitesimal magnetic fields are amplified up to their saturation energy $M$. 
Restarting DNS from the saturated state above $\Pm_c$, while gradually decreasing $Pm$ (dashed arrows), allows the dynamo branch to be tracked below its linear onset. The full \tcb{triangles} denote new results: they 
indicate the smallest energy $M_0$, of the optimal $\Bo$, found to trigger a self-excited dynamo. 
Time-series of magnetic and kinetic energies are shown in Fig.\ref{fig:TG1}\textit{(Right)} for $\Pm=0.44$. Here again, strong algebraic growth occurs first, followed by a short plateau. Exponential energy growth finally kicks in, as the optimal seed spontaneously evolves toward the saturated state corresponding to the (previously known) subcritical dynamo branch, driving larger, but slower, fluctuations in the kinetic energy. 
Note that the magnetic energy undergoes considerable amplification ($\sim 10^4$ times) during this process. On the other hand, the \tcb{empty triangles} in Fig.\ref{fig:TG1}(a) correspond to failed dynamos, suggesting that the frontier between the two (dynamo and non-dynamo) basins of attraction lies somewhere between the full and empty \tcb{triangles} lines. 

We re-emphasize that no assumptions were ever made about the structure or amplitude of the final state when searching for $\Bo$; furthermore, the structure of the minimal seed field (illustrated in Fig.\ref{fig:TG2}A) bears no topological resemblance with the saturated state 
(Fig.\ref{fig:TG2}D). This implies that identifying the former by means of DNS and continuation from the latter seems a hopeless endeavor. 
Instead, the minimal seed structure is found to be both very simple and localized in space, as well as being robust to changes in the initialization of the optimisation procedure. Although the nonconvexity of the optimisation problem considered, makes it impossible to guarantee that the optimal seed identified corresponds to a global extremum, repeated optimisations have reassuringly identified such magnetic loops. Remarkably, these seeds are located near one of the flow's stagnation points: indeed in such regions, the intense stretching of magnetic field lines ensures efficient growth of the magnetic energy \cite{Moffatt1978}, 
while localization ensures optimal expenditure of the energy budget $M_0$ \cite{Kerswell2018}. 

Finally, let us note that while the minimal energy $M_0$ required to trigger the subcritical dynamo seems to decrease slightly as $T$ increases, or that it changes slightly when maximising $\langle |\B(\bm{x},T)|^2 \rangle$ in place of its time-integrated quantity, the seed structure remains consistent in both cases. Due to the deterioration of the gradient estimates with larger optimisation times, or alternative objective functions in highly fluctuating regimes, these issues were not further investigated in the present study. Since we anticipate the minimal dynamo seed to be independent of the target time or the objective function if the former is sufficiently large, future improvement of the numerical methods used to estimate the gradient will be interesting for the purpose of accurate nonlinear stability analysis.

In this Letter we have shown how nonlinear optimisation, previously used to identify minimal disturbances in shear flows \cite{Pringle2010, Eaves2015,Kerswell2018}, is a numerically viable and flexible approach to systematically probe subcritical dynamo action in electrically conducting flows. The significance of this approach lies in its numerical feasibility, given the existence of highly parallel open source codes \cite{Burns2020}. It also lies in its here demonstrated ability to identify at the same time stable dynamo branches and their critical perturbations, without imposing symmetries, restricting time-dependence or making prior assumptions on the physical processes involved. Although we restricted our attention here to purely magnetic seeds, the same approach can readily identify optimal velocity, magnetic disturbances or a combination of both, with only minor modifications in the optimisation procedure. We \tcb{thus} argue that this approach represents a promising way of numerically obtaining elusive dynamo models, such as the strong branch of the geodynamo or the dynamo of Keplerian flows. Moreover, the possibility of placing additional constraints on the structure of the optimized disturbance can be exploited for experimental purposes, to design a way of kickstarting self-excited dynamos at sustainable energy cost in the laboratory.

\begin{acknowledgments}

\textbf{Acknowledgement -}  We wish to thank Didier Auroux for his helpful discussions regarding adjoint optimisation methods, and Jean-Marc Lacroix for his help in optimizing the installation of Dedalus. Y. P. thanks A. Miniussi for computing design assistance on the {\sc cubby} code. P.M.M. and F.M. acknowledge support from the French program `T-ERC' from Agence Nationale de la Recherche (Grant ANR-19-ERC7-0008-01). This work was also supported by the French government, through the UCAJEDI Investments in the Future project managed by the National Research Agency (ANR) under reference number ANR-15-IDEX-01. The authors are grateful to the OPAL infrastructure from Université Côte d'Azur, Université Côte d'Azur's Center for High-Performance Computing and GENCI computer facilities for providing resources and support.

\end{acknowledgments}

\bibliography{scibib}
\bibliographystyle{prl}

\end{document}